\documentclass[showpacs,prl,amsmath,amssymb,superscriptaddress,nobibnotes,twocolumn, preprintnumbers, bibnotes]
{revtex4-1}
\usepackage{graphicx}
\usepackage{endnotes}
\usepackage{bm}
\usepackage{epsfig}
\usepackage{amsmath}
\usepackage{bbm}
\usepackage{multirow}
\usepackage{color}
\usepackage{array}

\newcommand{\eg}{{\it e.g.}}

\newcommand{\Tstar}{$T^*$}

\begin{document}


\title{Nonsaturating large magnetoresistance in the high carrier density \\nonsymmorphic metal CrP}

\author{Q.~Niu}
\author{W.~C.~Yu}
\author{E.~I.~Paredes Aulestia}
\author{Y.~J.~Hu}
\author{Kwing~To~Lai}

\affiliation{Department of Physics, The Chinese University of Hong Kong, Shatin, New Territories, Hong Kong, China}

\author{H.~Kotegawa}
\author{E.~Matsuoka}
\author{H.~Sugawara}
\author{H.~Tou}
\affiliation{Department of Physics, Kobe University, Kobe 658-8530, Japan}

\author{D. Sun}
\author{F. F. Balakirev}
\affiliation{National High Magnetic Field Laboratory, Los Alamos National Laboratory, Los Alamos, New Mexico 87545, USA}

\author{Y.~Yanase}
\affiliation{Department of Physics, Kyoto University, Kyoto 606-8502, Japan}

\author{Swee~K.~Goh}
\email{skgoh@cuhk.edu.hk}
\affiliation{Department of Physics, The Chinese University of Hong Kong, Shatin, New Territories, Hong Kong, China}
\date{\today}


\begin{abstract}
The band structure of high carrier density metal CrP features an interesting crossing at the Y point of the Brillouin zone. The crossing, which is protected by the nonsymmorphic symmetry of the space group, results in a hybrid, semi-Dirac-like energy-momentum dispersion relation near Y. The linear energy-momentum dispersion relation along Y--$\Gamma$ is reminiscent of the observed band structure in several semimetallic extremely large magnetoresistance (XMR) materials. We have measured the transverse magnetoresistance of CrP up to 14~T at temperatures as low as $\sim$~16~mK. Our data reveal a nonsaturating, quadratic magnetoresistance as well as the behaviour of the so-called `turn-on' temperature in the temperature dependence of resistivity.
Despite the difference in the magnitude of the magnetoresistance and the fact that CrP is not a semimetal, these features are qualitatively similar to the observations reported for XMR materials. Thus, the high-field electrical transport studies of CrP offer the prospect of identifying the possible origin of the nonsaturating, quadratic magnetoresistance observed in a wide range of metals.

\end{abstract}


\maketitle

\section{Introduction}

The recent discovery of an extremely large magnetoresistance (XMR) in nonmagnetic topological semimetals has generated considerable attention \cite{Ali2014,Pletikosic2014,Wang2014,He2014,Ali2015,Liang2015,Wang2015,Shekhar2015,Luo2015,Lv2015,Wu2015,Huang2015,Tafti2016temperature,Tafti2016resistivity,Zeng2016,Sun2016,Wang2016,Wu2016,He2016,Hu2018}. At zero applied magnetic field, the electrical resistivity of XMR materials decreases with a decreasing temperature, following a typical metallic-like temperature dependence. Interestingly, at a modest magnetic field, the electrical resistivity begins to increase below a characteristic temperature, which is commonly called the `turn-on' temperature \cite{Ali2014,Ali2015}. Because of the low temperature upturn in the resistivity at high magnetic field, the magnetoresistance experiences a significant enhancement, giving rise to a remarkably large magnetoresistance. In WTe$_2$, for instance, the magnetoresistance can reach 452,700~\% at 14.7~T at 4.5~K \cite{Ali2014}.

The mechanism giving rise to the XMR behaviour is currently under debate \cite{Ali2014,Ali2015,Liang2015,Luo2015,Shekhar2015,Lv2015,Wu2015,Huang2015,Sun2016,Wu2016,Wang2015,He2016}. In particular, it is not clear if it is necessary to invoke the notion of nontrivial band topology to explain the origin of XMR \cite{Liang2015,Luo2015,Shekhar2015}. Indeed, several cases \cite{Ali2014,Ali2015,Pletikosic2014,Lv2015,Wu2015,Sun2016,Wu2016,Hu2018} exist which show that a simple two-band model is sufficient to quantitatively reproduce the extreme magnetoresistance. In XMR compound LaSb, angle-resolved photoemission spectroscopy (ARPES) failed to detect any nontrivial topology associated with the band structure \cite{Zeng2016}. Instead, it was argued that the nonsaturating magnetoresistance is simply due to a nearly perfect compensation of the holes and the electrons. To deepen the understanding of the XMR behaviour, it is important to study materials which exhibit similar behaviours.

The nonmagnetic metal CrP features an interesting band crossing at the Y point of the Brillouin zone \cite{Niu2017}. Along Y--$\Gamma$, the dispersion relation $E(k)$ is linear; along Y--S, the dispersion is parabolic. The crossing is four-fold degenerate, protected by the nonsymmorphic symmetry of the space group $Pnma$. The Fermi level is only $\sim$~47~meV above the crossing point \cite{Niu2017}, resulting in a small Fermi pocket centered at Y. If this were the only energy band, CrP would qualify to be a topological semimetal. However, CrP is a good metal that hosts multiple large Fermi surface sheets, in addition to the small pocket at Y. A detailed study of CrP has become even more urgent due to the recent discovery of the pressure-induced superconductivity in CrAs  \cite{Wu2014,Kotegawa2014}, which is isostructural to CrP. With the application of a moderate pressure of $\sim$7~kbar, the helimagnetic state in CrAs can be suppressed, accompanied by the birth of the superconductivity. The band structure in the paramagnetic state obtained by density functional theory is similar in two compounds \cite{Niu2017}. In particular, the nontrivial band crossing at the Y point is preserved in CrAs, but the crossing point is even closer to the Fermi level, giving rise to a tiny Fermi pocket which is believed to be responsible for new Landau levels, and an unusual quasilinear magnetoresistance at low temperatures \cite{Niu2017}.

In this manuscript, we report an extensive dataset of the transverse magnetoresistance of a highly pure CrP single crystal as a function of temperature, magnetic field strength and field angle, down to $\sim$16~mK and up to 14~T. Despite the obvious difference in the carrier density from typical topological semimetals, CrP exhibits many features that are characteristic of XMR materials, including the `turn-on' temperature and a nonsaturating magnetoresistance, although the magnitude of the magnetoresistance is smaller in CrP. In addition, high-quality Shubnikov-de Haas oscillations can be resolved from our data, allowing the construction of the Fermi surface. Hence, our detailed study of CrP serves two purposes: it provides a large carrier density reference for comparison with typical semimetals showing XMR, and simultaneously it is an important reference compound to understand the normal state of the pressure-induced superconducting state in CrAs.
\begin{figure}[!t]\centering
      \resizebox{9cm}{!}{
              \includegraphics{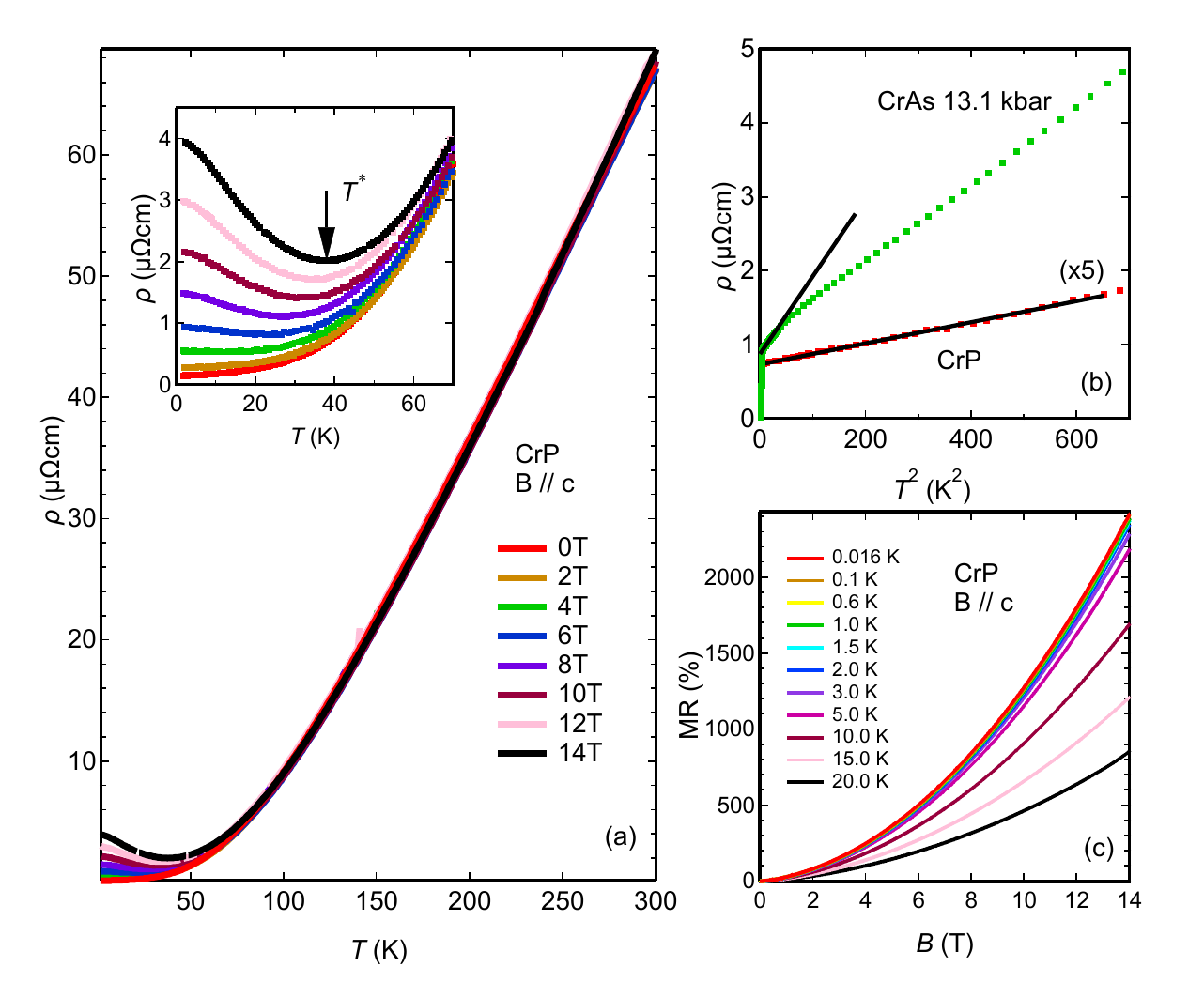}}                				
              \caption{\label{fig1} (a) Temperature dependence of electrical resistivity of CrP at 0,~2,~4,~6,~8,~10,~12,~14~T. The inset shows the low temperature upturns. The black solid arrow indicates $T^*$ at 14~T. (b) The zero-field resistivity of CrP versus $T^2$ compared with that of CrAs at 13.1~kbar. The resistivity of CrP is multiplied by 5 times. The black solid lines show the $T^2$ region. (c) The MR of CrP at different temperatures. The magnetic field is applied along the $c$-axis.}
\end{figure}
\section{Experimental details}

Single crystals of CrP were synthesized by the chemical vapour transport method \cite{Nozue1999}, utilizing iodine as the transport agent. The charge zone was set to 900 $^\circ$C and the growth zone was kept at 800 $^\circ$C for 2 weeks. The electrical resistivity was measured with the conventional four-probe configuration with the current flowing along the crystallographic $a$ axis. A dilution refrigerator (BlueFors) equipped with a 14~T superconducting magnet was used to cool down the sample to $\sim$16~mK. The sample was placed on a rotator at the center of the magnet. The rotation axis was parallel to the crystallographic $a$ axis, and hence the magnetic field is always perpendicular to the current direction. The magnetic field angle was monitored by a Hall sensor glued on the rotator platform. Reference data of CrAs at 13.1~kbar were collected as described in Ref. \cite{Niu2017}. The electronic structure was calculated using the all-electron full-potential linearized augmented plane-wave code WIEN2k \cite{Schwarz2003}. The generalized gradient approximation (GGA) of Perdew, Burke and Ernzerhof (PBE) \cite{Perdew1997} was employed for the exchange-correlation potential. Experimental lattice constants at 4.2~K \cite{Selte1975} were used in the calculation and internal structure optimization was performed. The muffin-tin radii were set to 1.96~a.u. for the P atoms and 2.24~a.u. for the Cr atoms. R$_{\rm MT}^{\rm min}$K$_{\rm max}$=8 and a $k$-point mesh of 10000 in the first Brillouin zone were used. The quantum oscillation frequencies were extracted using the Supercell K-space Extremal Area Finder (SKEAF) code \cite{Rourke2012}.

\section{Results and Discussion}

Figure~\ref{fig1}(a) shows the temperature ($T$) dependence of electrical resistivity ($\rho$) in CrP at different magnetic field. The zero field $\rho(T)$ exhibits a typical metallic behaviour. The small residual resistivity ($\rho_0$) of 0.15~$\mu\Omega$cm and the large residual resistance ratio (RRR) of 455 indicate superior crystal quality. As magnetic field increases, $\rho(T)$ exhibits a more complicated behaviour at low temperatures, as shown in the inset of Fig.~\ref{fig1}(a). Above 6~T, it begins to show an upturn below $T^*\sim25$~K . This is the so-called `turn-on' behaviour, which has been observed in all XMR materials studied so far \cite{Ali2014,Ali2015,Wang2014,Shekhar2015,Tafti2016resistivity,Tafti2016temperature,Sun2016,Wang2016,Wu2016,Zeng2016,Wang2015,Hu2018}. \Tstar\ increases with an increasing magnetic field, and it reaches $\approx$~38~K at 14~T, as indicated by the arrow in the inset of Fig.~\ref{fig1}(a).
As shown in Fig.~\ref{fig1}(b), the zero field resistivity can be described by $\rho(T)=\rho_0+AT^2$ below 25~K, with an extremely small $A=0.29~{\rm n}\Omega$cmK$^{-2}$. For comparison, the data of CrAs at 13.1~kbar is plotted, which shows a very narrow $T^2$ region with a $A$-coefficient which is $\approx37$ times larger. The large difference represents a much stronger electron-electron correlation in CrAs at 13.1~kbar, which may be attributed to the magnetic fluctuation or the proximity to a quantum critical point (QCP) \cite{Kotegawa2015, Matsuda2018}. The Kadowaki-Woods ratio $A/\gamma^2$ in CrP is calculated to be 3.7~$\mu\Omega$cm.mol$^2$K$^2$J$^{-2}$ (where $\gamma=8.86$~mJmol$^{-1}$K$^{-2}$ from our specific heat data \cite{SUPP}). This value is nearly 10 times higher than that of transition metals, but about 2.5 times lower than that of typical heavy fermions \cite{Kadowaki1986, Jacko2009}. Therefore, CrP can be regarded as a moderately correlated electron system. In Fig.~\ref{fig1}(c), the magnetoresistance (MR), defined as $[\rho(B)-\rho(0)]/\rho(0)\times100\%$, is plotted against the magnetic field ($B$) for temperatures spanning three orders of magnitude from less than $20$~mK to 20~K. The MR at all temperatures is almost proportional to $B^2$, and shows no sign of saturation at 14~T. The overall magnitude of MR increases with decreasing temperature, reaching a value of about 2500\% at 16~mK and 14~T. Additional high-field MR data up to 58 T also show no saturation of MR, as presented in the Supplemental Material \cite{SUPP}.
\begin{figure}[!t]\centering
       \resizebox{9cm}{!}{
              \includegraphics{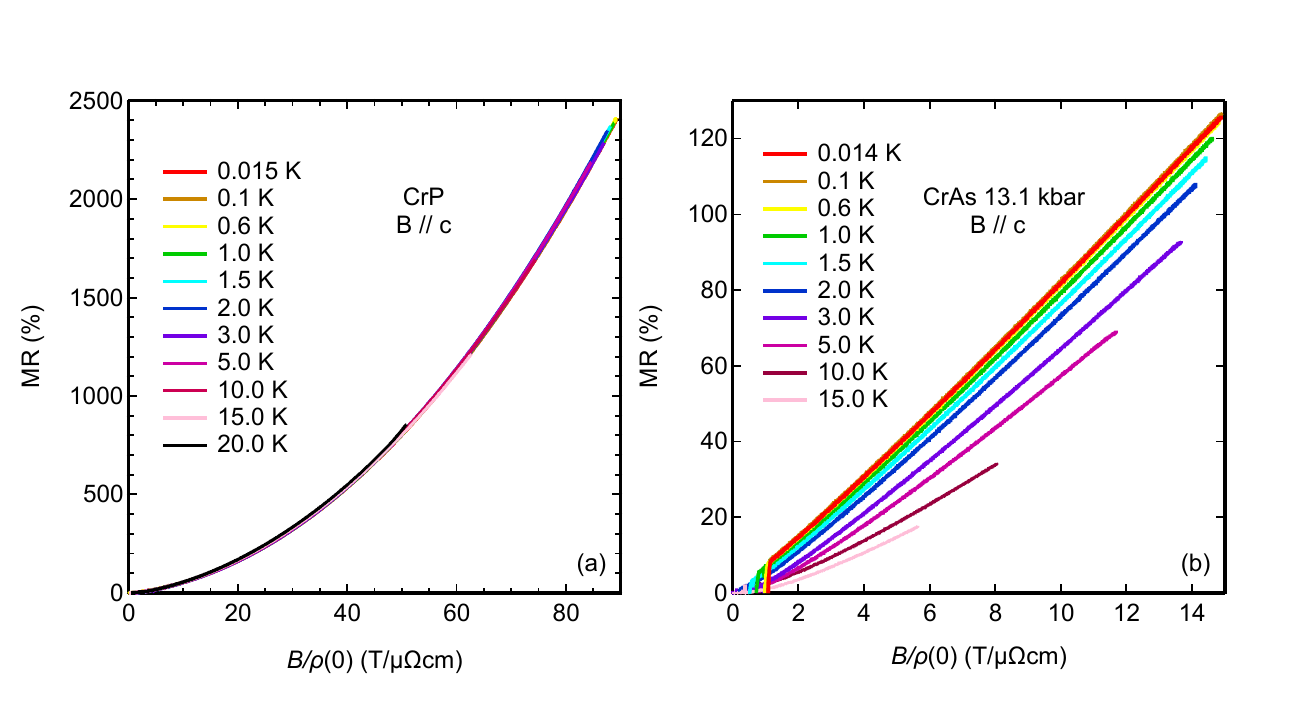}}                				
              \caption{\label{fig2} The Kohler plot of (a) CrP compared with that of (b) CrAs at 13.1~kbar. The magnetic field is applied along the $c$-axis. The magnetoresistance of CrAs at 13.1~kbar are taken from Ref.~\cite{Niu2017}. }
\end{figure}

In Fig.~\ref{fig2}(a), the MR data are replotted against $B/\rho(0)$. This is the so-called Kohler plot, and it is clear that all MR curves at different temperatures can be satisfactorily scaled onto a single universal curve, highlighting a universal scattering mechanism. Phenomenologically, the observation of the Kohler's scaling ${\rm MR}\propto(B/\rho(0))^2$ naturally leads to the turn-on behaviour described above, provided that the crystal is sufficiently pure \cite{Wang2015, Sun2016}. This treatment rules out exotic field-induced scattering mechanism \cite{Khveshchenko2001,Kopelevich2006}, \eg\ a field-induced metal-insulator transition, and the turn-on behaviour is simply due to the extraordinarily high sample mobility \cite{Wang2015, Sun2016}. As a comparison, the MR becomes more unconventional upon moving towards the CrAs side. In CrAs at 13.1~kbar, the MR is no longer quadratic in field \cite{Niu2017}, accompanied by a strongly violated Kohler's rule, as evidenced in Fig.~\ref{fig2}(b). 

\begin{figure}[!t]\centering
       \resizebox{9cm}{!}{
              \includegraphics{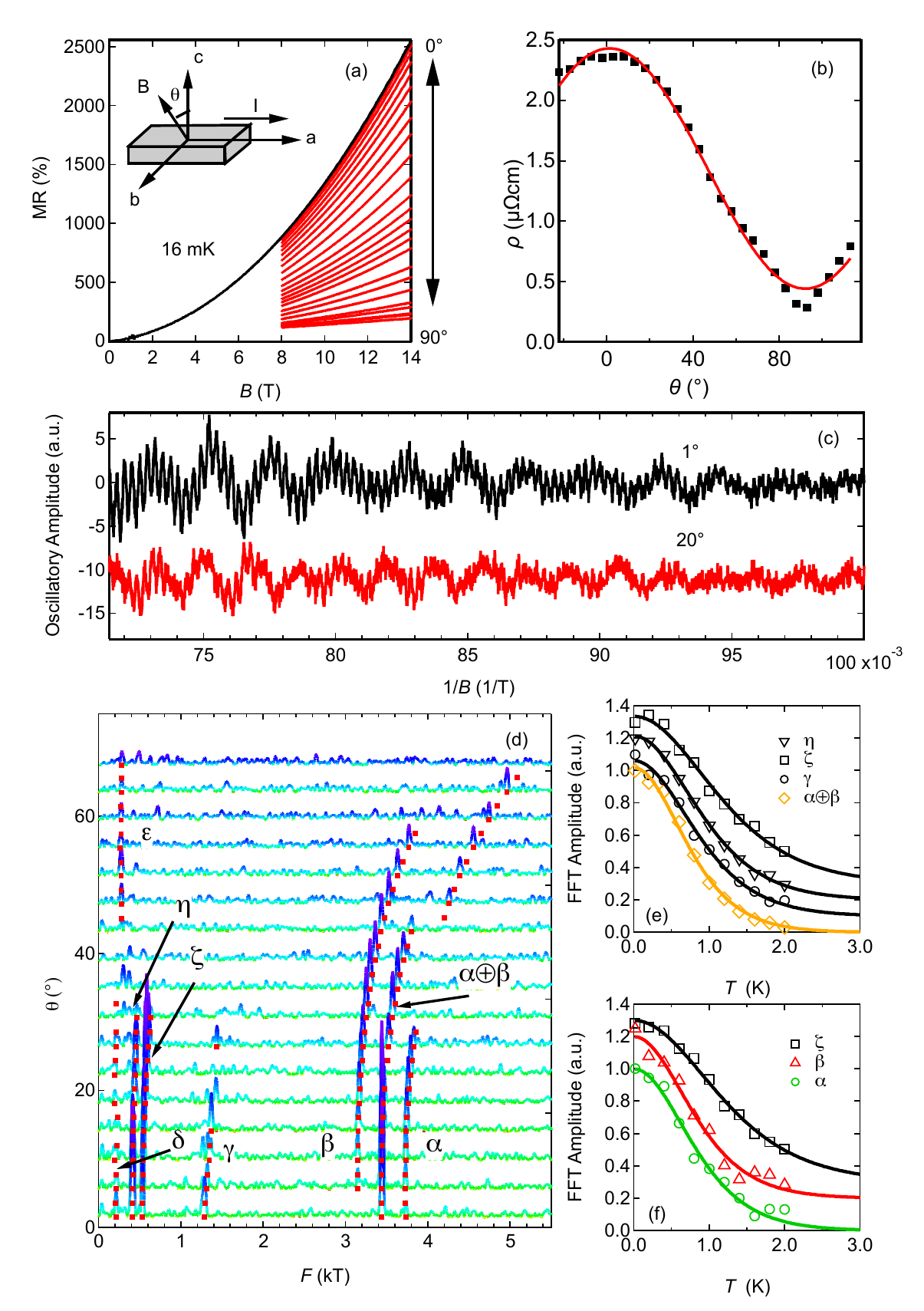}}                				
              \caption{\label{fig3} (a) The MR of CrP measured at 16~mK between 8~T and 14~T at different angles. The inset is a schematic diagram showing the definition of the rotation angle $\theta$. (b) The resistivity of CrP at 14~T and 16~mK as a function of $\theta$. The black solid symbols are experimental results and the red solid line is the fit to a sinusoidal function.   (c) Oscillatory data at $\theta=1^\circ$ and $\theta=20^\circ$ at 16~mK with the background subtracted. (d) FFT spectra of the SdH oscillations at different $\theta$ at 16~mK. The red dots indicate the observed peak positions at different angles. (e, f) Temperature dependence of SdH amplitudes (symbols) at $\theta=1^\circ$ and $\theta=20^\circ$, respectively, analyzed with the Lifshitz-Kosevich formula (solid curves).  }
\end{figure}

At 16~mK, we measured the MR of CrP from 8~T to 14~T at different $\theta$, where $\theta$ is the angle between the $c$-axis and the magnetic field direction as shown in the inset of Fig.~\ref{fig3}(a). As the magnetic field rotates, the MR maintains the $B^2$ dependence, although the magnitude decreases from about 2500\% at $c$-axis to about 200\% at $b$-axis at 14~T, as shown in Fig.~\ref{fig3}(a). Figure~\ref{fig3}(b) displays the angular dependence of the resistivity at 14~T and 16~mK (solid symbols), which roughly follows $(\cos^2(\theta)+\gamma_{\rm ani}^{-2}\sin^2(\theta))$ expected for an anisotropic Fermi surface, giving rise to an anisotropy $\gamma_{\rm ani}\sim13$.  Our experimental configuration ensures a roughly constant Lorentz force. Therefore, the anisotropy most likely reflects the anisotropy of the underlying electronic structure. We note that the rather flat maximum near $0^\circ$ and the sharp, pointy minimum near $90^\circ$ may come from the multiband nature of the system, which contains several Fermi surface sheets with complicated shapes (see below).

To confirm the high carrier density nature of CrP, we calculated and measured the Fermi surface of CrP via density functional theory (DFT) and Shubnikov-de Haas (SdH) oscillations, respectively. With the large magnetoresistance background removed, clear SdH signals can be observed. Figure~\ref{fig3}(c) shows the SdH oscillations of CrP at 16~mK for $\theta=1^\circ$ and $\theta=20^\circ$. We analyzed such field sweeps at different $\theta$ with fast Fourier transform (FFT). Figure~\ref{fig3}(d) displays the outcome of the analysis, showing the variation of the SdH amplitude against the SdH frequency and $\theta$. The SdH frequencies vary smoothly with $\theta$, resulting in branches that are labelled by Greek letters. From the FFT spectra, three groups of SdH branches can be seen: one group with frequencies larger than 3~kT, one group with frequencies less than 1~kT, and a single branch at around 1.25~kT. The high-frequency group exhibits a strong angular dependence, while the other groups are almost independent of $\theta$. At $\theta=1^\circ$, we obtain peaks at 3.44~kT, 1.29~kT and about 0.5~kT, which are consistent with previous de Haas-van Alphen (dHvA) results \cite{Nozue1999}. The key difference between our work and the dHvA work is that the authors of Ref.~\cite{Nozue1999} attributed the three branches with minima of $\sim$3.1~kT to 3.7~kT near $B\parallel c$ to ellipsoidal Fermi surfaces. However, our calculations indicate that two of them come from quasi-2D Fermi surfaces and the middle branch is a breakdown orbit (see below). Additionally, we observed two new, small frequencies at 210~T and 410~T. 

From the temperature dependence of the SdH amplitudes, we can extract the cyclotron effective masses ($m^*$) associated with the frequencies with sufficiently strong amplitudes using a standard Lifshitz-Kosevich prescription \cite{Shoenbergbook}. Figures~\ref{fig3}(e) and (f) display the resultant fits of the SdH amplitudes at $\theta=1^\circ$ and $\theta=20^\circ$, respectively.  With these analyses, combined with the detailed knowledge of the cyclotron orbits to be discussed below, we are able to obtain $m^*$ associated with most frequencies. As tabulated in Table~\ref{Table1}, most $m^*$ have a value of the order of 2$m_e$, consistent with the claim that CrP is a moderately correlated metal.


\begin{table}[!t]
\centering
\caption{The cyclotron effective mass ($m^*)$ associated with experimentally observed Shubnikov-de Haas frequencies, and their corresponding bands from DFT calculations. The hyphens (--) indicate no observation in the experiment.}
\label{Table1}
\begin{tabular}{ccc||ccc}
\hline
 Branch 			& $m^*/m_e$ 			& Bands 			& Branch 			& $m^*/m_e$ 		& Bands \\ 
 \hline
$\alpha$  			&  2.08(8)         			& 75/76   			& $\zeta$			&1.46(4)			&77/78\\
$\beta$ 			&  2.08(10) 			& 73/74  			& $\eta$			&1.84(2)			&77/78 \\
$\alpha\oplus\beta$ 	& 2.29(6) 				& 73/74 \& 75/76    	&  $\varepsilon$	&     -- 			&79/80\\
$\gamma$		&1.94(5)				&77/78 			& $\delta$			&	--			&79/80 \\  
\hline
\end{tabular}
\end{table}

Since the SdH frequency ($F$) is related to the area of the Fermi surface extremal orbit ($S_F$) via the Onsager's relation $F=\frac{\hbar}{2\pi e}S_F$, the full angular dependence $F(\theta)$ enables a comparison with the calculated Fermi surface. Band structure calculations with spin-orbit coupling (SOC) included show that four pairs of bands cross the Fermi energy. We extract the size of extremal orbits, and hence the associated quantum oscillation frequencies, from the calculated Fermi surfaces. The extracted spin-up and spin-down frequencies from each pair are practically identical, indicating an exceedingly weak energy splitting. In other words, SOC is negligible, consistent with the expectation for Cr-based electronic systems. 

\begin{figure}[!t]\centering
       \resizebox{8cm}{!}{
              \includegraphics{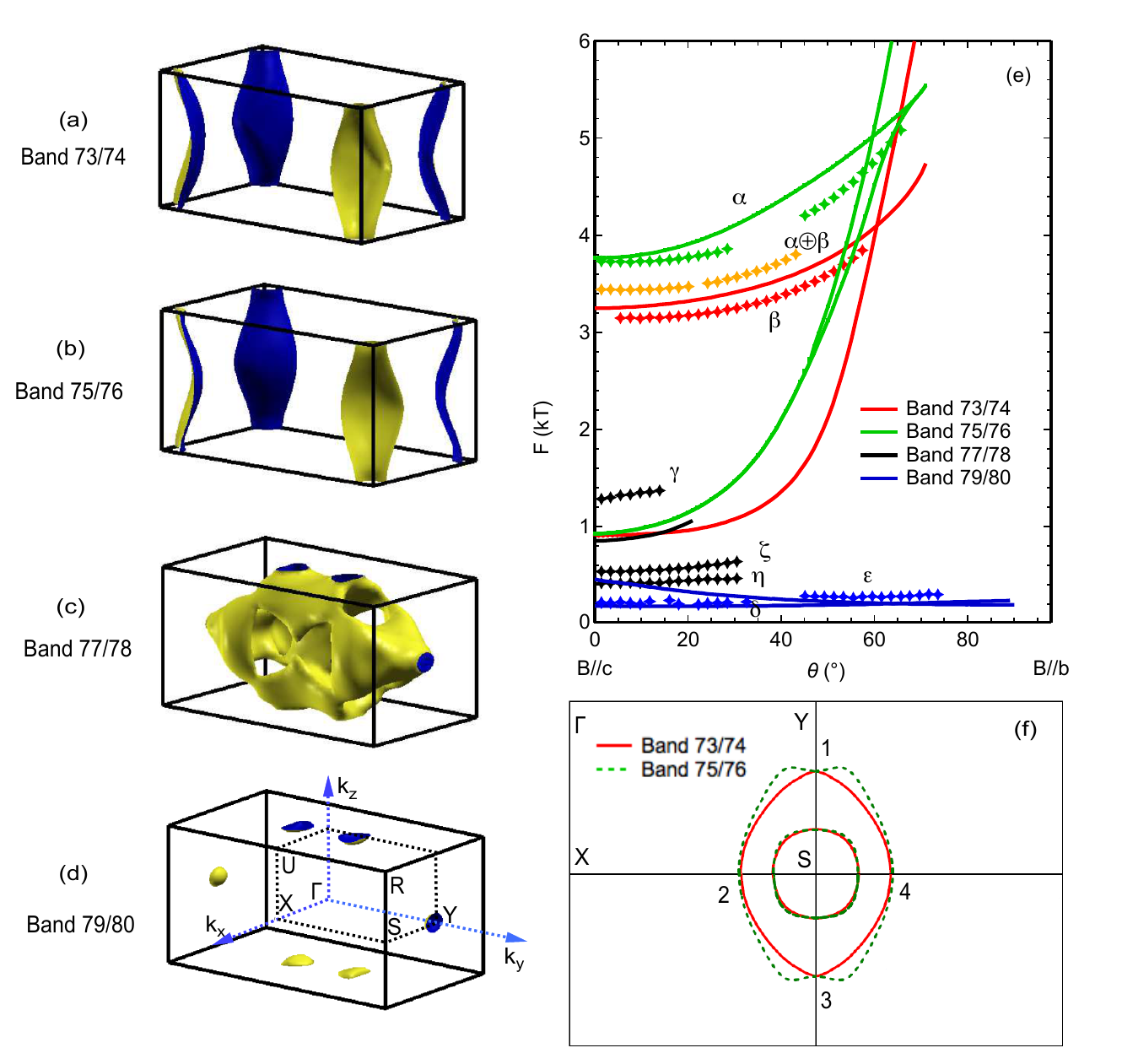}}                				
              \caption{\label{fig4} (a)--(d) Fermi surfaces of CrP. Bands 73/74 and 75/76 are hole-like, and Bands 77/78 and 79/80 are electron-like. The high symmetry points are labelled in (d). (e) Comparison of the angular dependence of calculated SdH frequencies (solid lines) and the experimental data (symbols). (f) The projection of the calculated Fermi surfaces from Bands 73/74 and 75/76 on the $k_x$-$k_y$ plane.
              }
\end{figure}

Figures~\ref{fig4}(a)--(d) displays the calculated Fermi surfaces from Bands 73/74, 75/76, 77/78 and 79/80. The usage of pairs of numbers is due to spin degrees of freedom. In Fig.~\ref{fig4}(e), the extracted frequencies from these bands are shown in lines \cite{SUPP}, with the experimental results displayed as symbols.
The qualitative agreement between experimental data and the calculated Fermi surfaces for these bands can be seen.
Band 73/74 and Band 75/76 give rise to quasi-2D hole cylinders centered at S in the Brillouin zone. In Fig.~\ref{fig4}(e), the calculated ``belly" frequencies (red and green solid lines) of these cylinders match well with the $\beta$ and $\alpha$ branches respectively. However, an additional branch labelled as ``$\alpha\oplus\beta$" is experimentally observed in between $\alpha$ and $\beta$ branches, with the value roughly equal to $(F_\alpha+F_\beta)/2$. The appearance of this branch can be understood as the ``magnetic breakdown" involving $\alpha$ and $\beta$ orbits. Figure~\ref{fig4}(f) shows the calculated extremal orbits from Bands 73/74 and 75/76 projected on to the $k_x$-$k_y$ plane when the field is along $k_z$. The inner orbits correspond to the neck orbits, while the larger, outer orbits are $\beta$ and $\alpha$, which are the belly orbits of the quasi-2D Fermi surfaces. It can be seen that $\beta$ and $\alpha$ touch at points labelled as ``1" and ``3" along the S--Y direction. These degeneracies are protected by the nonsymmophic symmetry. Along the S--X direction $\beta$ and $\alpha$ almost meet at ``2" and ``4". In fact, in the complete absence of the spin-orbit coupling, the orbits would also meet at ``2" and ``4".
When the magnetic field is tilted towards the $b$-axis, the extremal orbits of $\alpha$ and $\beta$ gradually expands but they remain degenerate at ``1" and ``3" along S--Y.
Hence, the quasiparticle can always hop between $\beta$ and $\alpha$ at ``1" and ``3", resulting in a breakdown orbit with frequency $(F_\alpha+F_\beta)/2$, as observed. The bandstructure of CrP possesses a novel mechanism to realize the magnetic breakdown.

Band 77/78 is complicated with many closed orbits oriented along different directions. Unfortunately, we are unable to uniquely match the observed and the calculated frequencies. Band 79/80 contains two small Fermi pockets as shown in Fig.~\ref{fig4}(d), which give two low-frequency peaks of $\sim$200~T. Experimentally, $\delta$ and $\varepsilon$ agree with the calculated frequencies. However, these peaks are weak and they disappear very quickly at higher temperatures, making it impossible to reliably obtain the effective masses. Nevertheless, our SdH data broadly agree with the calculated Fermi surfaces.

Using a simple two-band model, a non-saturating, quadratic MR can be expected if the electron carrier density ($n_e$) is equal to the hole carrier density ($n_h$) \cite{Singletonbook}. In CrP, the calculated carrier densities are $n_e=3.68\times10^{21}$~cm$^{-3}$ and $n_h=3.54\times10^{21}$~cm$^{-3}$, resulting in the ratio $n_e/n_h=1.04$. On the other hand, simultaneous fitting of $\rho_{xx}(B)$ and $\rho_{yx}$(B) using the two-band model under the assumption of $n_e/n_h=1$ gives $n_e=n_h=3.1\times10^{21}$~cm$^{-3}$, which is in good agreement with the DFT estimates.
Thus, CrP is nearly perfectly compensated, and the observation of a non-saturating, quadratic MR (Figs.~\ref{fig1}(c) and S2) is consistent with the two-band model. However, we note that the carrier densities in CrP are nearly more than two orders of magnitude larger than that of most topological semimetals showing similar XMR behaviors. Therefore, our work shows that XMR behaviour is not limited to low-carrier density topological semimetals. Furthermore, despite its semiclassical origin, the two-band model can be applied to the present case with a large carrier density and complicated Fermi surface shapes.

\section{Summary}

In summary, we have measured the magnetoresistance of CrP, which shows features similar to many topological semimetals. Our SdH oscillations and DFT calculations uncover a novel magnetic breakdown mechanism originated from degeneracies protected by the nonsymmorphic symmetry of the space group, and establish the applicability of the two-band model in a high-carrier density system with complex Fermi surfaces. Recently, a measurement of the angular dependence of the upper critical field of CrAs under pressure suggests the presence of spin-triplet superconductivity  \cite{Guo2017}. A theoretical calculation based on crystalline symmetries of $Pnma$ space group predicts that CrAs is a promising candidate to realize topological nonsymmorphic crystalline superconductivity \cite{Daido2018}. Since the Fermi surfaces for both CrAs and CrP are very similar \cite{Niu2017}, our magnetotransport data serve to establish a useful benchmark for understanding the normal state of CrAs. 

\begin{acknowledgments}
We acknowledge financial support from Research Grants Council of Hong Kong (GRF/14300418, GRF/14301316), CUHK Direct Grant (4053223, 4053299), Grants-in-Aid for Scientific Research (KAKENHI) (15H05882, 15H05884, 15H05885, 15K05164, 15H05745, 15H03689, 18H04321, 18H01178, 18H04225, and 18H05227). The National High Magnetic Field Lab Pulsed Field facility is supported by the National Science foundation under cooperative Grant Nos. DMR-1157490 and DMR-1644779, the U.S. DOE, and the State of Florida.

\end{acknowledgments}




\end{document}